# Improving 5G/B5G Network Performance with RFID-Enabled Resource Management Systems

**Stella N. Arinze [1], Halima I. Kure [2] and Augustine O. Nwajana [3]**

[1] Dept. of Electrical and Electronic Engineering, Enugu State University of Science and Technology, Enugu, PMB 01660, Nigeria; ndidi.arinze@esut.edu.ng
[2] School of Architecture, Computing and Engineering, University of East London, London E16 2RD, UK; Hkure2@uel.ac.uk
[3] School of Engineering, University of Greenwich, Chatham Maritime, Kent ME4 4TB, UK; a.o.nwajana@greenwich.ac.uk

**Abstract**
In the rapidly evolving landscape of 5G and B5G (beyond 5G) networks, efficient resource optimization is critical to addressing the escalating demands for high-speed, low-latency, and energy-efficient communication. This study explores the integration of Radio Frequency Identification (RFID) technology as a novel approach to enhance resource management in 5G/B5G networks. The motivation behind this research lies in overcoming persistent challenges such as spectrum congestion, high latency, and inefficient load balancing, which impede the performance of traditional resource allocation methods. To achieve this, RFID tags were embedded in critical network components, including user devices, base stations, and Internet of Things (IoT) nodes, enabling the collection of real-time data on device status, location, and resource utilization. RFID readers strategically placed across the network continuously captured this data, which was processed by a centralized controller using a custom-designed optimization algorithm. This algorithm dynamically managed key network resources, including spectrum allocation, load balancing, and energy consumption, ensuring efficient operation under varying network conditions. Simulations were conducted to evaluate the performance of the RFID-based model against traditional 4G dynamic resource allocation techniques. The results demonstrated substantial improvements in key performance metrics. The proposed system achieved a 25% increase in spectrum utilization, a 30% reduction in average latency, a 15% boost in network throughput, and a 20% decrease in overall energy consumption. These gains highlight the effectiveness of the RFID-based optimization model in meeting the stringent performance requirements of 5G networks, particularly in high-density deployments. This study provides a scalable, cost-effective solution for optimizing resource management in 5G/B5G and lays the groundwork for future advancements in 6G networks. By leveraging real-time data and intelligent resource allocation, the proposed model addresses critical challenges in modern communication systems, ensuring enhanced network efficiency, reliability, and sustainability.

**Keywords: Radiofrequency identification; resource optimization; 5G/6G network; network performance**

## 1. Introduction

The demand for fast, reliable, and low-latency connectivity is rapidly increasing, driven by advancements such as IoT, autonomous systems, AR, and VR. 5G networks are essential to support these innovations, offering ultra-reliable low-latency communication (URLLC) and massive machine-type communications (mMTC). However, deploying and managing 5G infrastructure poses challenges, particularly in high-density urban areas. Traditional 4G resource management techniques are insufficient for 5G's complexities, leading to issues like spectrum congestion, energy inefficiency, and service delays. Optimizing resources in 5G is therefore critical to enabling scalable, efficient, and reliable connectivity for applications like smart cities and autonomous systems. Research has explored innovative strategies for 5G resource management, with RFID technology emerging as a promising solution. RFID provides real-time data on device location, movement, and network conditions, enabling intelligent resource allocation. Widely applied in inventory management and logistics, RFID has demonstrated its potential to address 5G's challenges by integrating tags into user devices, base stations, and IoT nodes.

Studies highlighted successful 5G resource management approaches, such as dynamic allocation using Software-Defined Networking (SDN), which improves efficiency and reduces congestion [1]. Deep learning algorithms enhance spectrum utilization, throughput, and energy efficiency [2], while RFID-integrated systems prioritize high-priority tasks and ensure reliable transmissions for critical applications like autonomous driving [3]. AI-based strategies further optimize energy consumption while maintaining performance [4,5]. Combining RFID with machine learning offers significant potential for spectrum optimization and load balancing. RFID also supports energy-efficient operations by enabling adaptive power management, reducing waste, and lowering costs. This study examines RFID's integration into 5G resource management, using real-time data to optimize bandwidth, power, and load balancing. RFID-based optimization enhances spectrum use, reduces latency, and minimizes energy consumption, paving the way for efficient, sustainable 5G networks and laying a foundation for 6G advancements. By integrating RFID with IoT, machine learning, and edge computing, this framework addresses current challenges while preparing for future demands.

## 2. Theory of Work
### 2.1 5G Network Resource Management and Optimization Challenges

5G networks, the next generation of wireless communication technology, deliver ultra-fast speeds, low latency, and massive connectivity, enabling advanced applications like IoT, autonomous systems, and smart cities. As shown in Figure 1, 5G architecture incorporates technologies such as millimeter waves, massive MIMO, and network slicing to optimize performance and efficiency. Designed to handle high bandwidth, ultra-low latency, and extensive device connectivity, 5G supports applications like autonomous vehicles, remote healthcare, and smart cities [6]. Effective resource management is a critical challenge due to increased device density, diverse QoS requirements, and dynamic user demands in urban areas. Unlike 4G, which relies on reactive resource allocation, 5G requires proactive, real-time strategies to address fluctuating network conditions. Traditional methods, including fixed spectrum allocation and proportional fair scheduling, cannot meet 5G's complexity, necessitating intelligent, data-driven approaches for resource optimization [7–9]. With the capacity to support up to one million devices per square kilometer, 5G demands innovative solutions to prevent congestion and latency issues. Real-time strategies are crucial for enabling ultra-reliable low-latency communication (URLLC), enhanced mobile broadband (eMBB), and massive machine-type communication (mMTC) [10]. Technologies like RFID offer a promising solution by providing real-time network data to optimize resource allocation, ensuring efficient and scalable 5G operations.

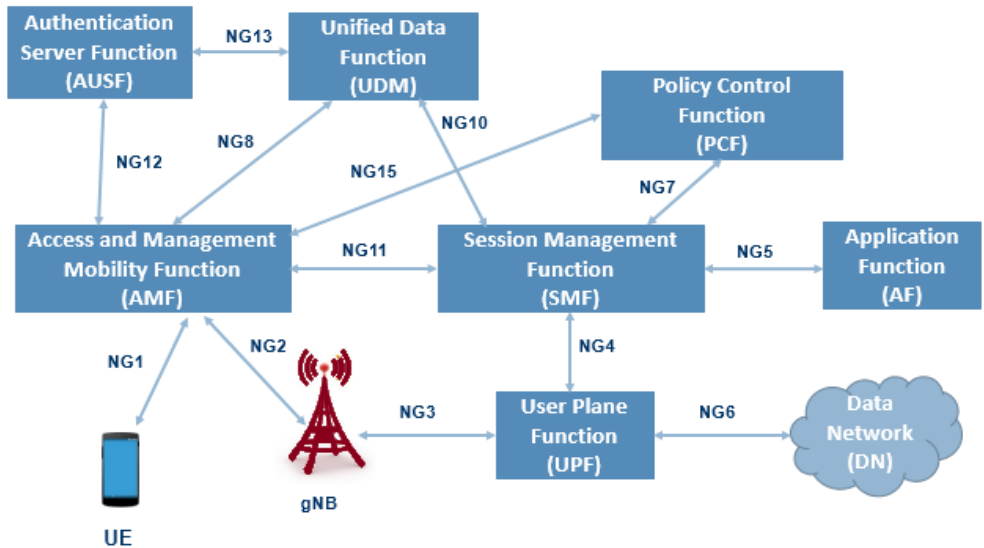

**Figure 1.** 5G Network Architecture

## 2.2 RFID Technology and Its Application in Network Resource Management

Radio Frequency Identification (RFID) is a wireless technology that uses electromagnetic fields to identify and track objects, animals, or people via embedded tags. As shown in **Figure 2**, RFID systems comprise tags, readers, and a backend system that processes real-time data on item location and movement, enhancing operational efficiency through automation. Initially used in inventory tracking and asset monitoring, RFID now plays a critical role in 5G networks by enabling real-time data collection on device location, usage patterns, and network status [11].Embedding RFID tags in network components such as devices, base stations, and IoT nodes allows readers across the network to gather data, which a centralized controller processes to optimize resource allocation dynamically. This improves load balancing, spectrum allocation, and overall network situational awareness [12]. For example, in dense urban areas with fluctuating connectivity demands, RFID data facilitates efficient resource distribution to high-demand zones. Its scalability makes RFID ideal for 5G networks, allowing for seamless expansion by adding more tags or readers as needed. Additionally, RFID's affordability supports cost-effective, large-scale deployment, reducing operational and maintenance expenses over time.

RFID also enhances network security by tracking devices and network elements, enabling the detection of unauthorized access and safeguarding sensitive data. Authors in [13] emphasized that real-time RFID data improves resource allocation efficiency, especially during peak traffic, ensuring uninterrupted critical services. Furthermore, researchers in [14] highlighted RFID's role in energy optimization by allocating resources only where needed, reducing power consumption across base stations. By providing enhanced visibility, RFID enables network operators to anticipate congestion, balance loads, and maintain 5G performance under demanding conditions.

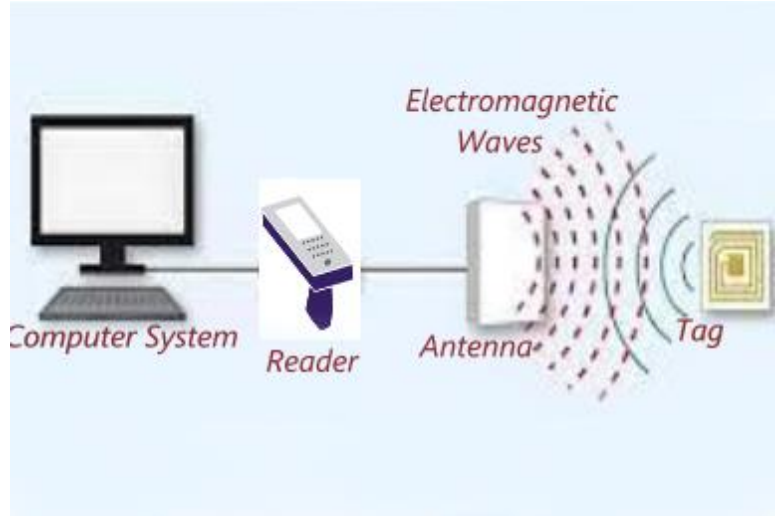

**Figure 2.** Basic RFID System

## 2.3 Machine Learning and RFID-Driven Predictive Resource Optimization in 5G

Machine learning (ML), a branch of artificial intelligence, enables systems to learn from data and improve performance over time without explicit programming. By leveraging algorithms and statistical models, ML powers applications such as recommendation systems and autonomous vehicles, driving innovation by automating complex tasks and extracting insights from large datasets. In 5G networks, ML transforms RFID data into actionable insights for resource management. By analyzing historical and real-time RFID data, ML algorithms predict network conditions, identify user behavior patterns, and allocate resources proactively [15]. Techniques like reinforcement learning, neural networks, and decision trees process large volumes of RFID data to optimize bandwidth, reduce latency, and meet QoS requirements. For example, reinforcement learning adapts resource allocation strategies based on feedback, ensuring balanced loads and efficient spectrum use [16]. Supervised learning models, including support vector machines (SVM) and random forests, classify network traffic and forecast congestion based on RFID data, enabling rapid adjustments to shifts in user patterns or device density [17]. This approach is especially effective in IoT applications, where sudden device increases could overwhelm resources [18]. Integrating ML with RFID data enhances 5G network management by enabling real-time, adaptive decision-making [19]. This improves resilience during peak usage, reduces downtime, and ensures a superior user experience. Additionally, ML-driven predictive capabilities offer scalable, cost-effective solutions that adapt as networks grow, with continuous learning minimizing risks such as congestion, latency, and security breaches. As networks evolve toward 6G, the combination of ML and RFID will become increasingly essential for resource optimization and reliability [20].

## 3. Materials and Methods

An RFID-based resource optimization system for 5G networks was developed using Arduino Mega, ESP32, a virtual terminal in place of the MFRC522, and NetSim. This architecture enables efficient data acquisition, processing, and optimization in a simulated environment. The system's key stages include initialization, data acquisition and processing, running the optimization algorithm, network condition checks, and integration of Proteus and NetSim. The block diagram of the system is shown in Figure 3.

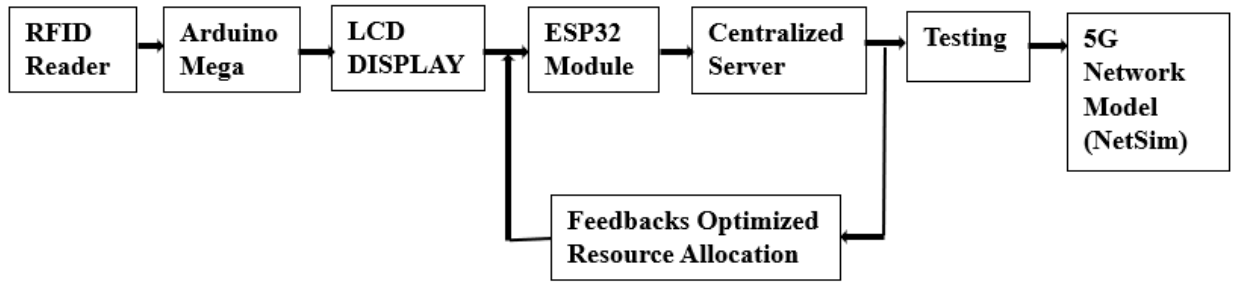

**Figure 3. Block Diagram of RFID-based Resource 0ptimization System for 5G networks**

The hardware components in Figure 3 were chosen for their compatibility with IoT-based 5G systems. The Arduino Mega provides robust processing, while the ESP32 ensures wireless communication for IoT integration. An LCD and LED offer real-time feedback, and the virtual terminal emulates an RFID reader for flexible testing. The RFID Reader initiates data acquisition, transmitting tag data to the Arduino Mega for validation and processing. The detected tag and its access status are displayed on a 16x2 LCD, and access is indicated via an LED. Processed data is sent from the Arduino Mega to the ESP32, which acts as a wireless gateway, transmitting the data to a centralized server. Optimization algorithms on the server dynamically allocate resources and return the updated allocation data to the ESP32, which updates the system status on the LCD. Section 3.4 validates the efficiency of this feedback mechanism. The integration of Proteus and NetSim, explained in Section 3.5, enables bidirectional communication between the hardware simulation (Proteus) and the 5G network model (NetSim). The ESP32 plays a key role as a gateway, facilitating data exchange between the two platforms. Sections 3.1–3.5 provide detailed descriptions of the hardware connections, optimization processes, and integration mechanisms.

### 3.1 System Initialization

The hardware schematic, developed in Proteus, integrates the Arduino Mega (ATmega2560) as the primary microcontroller for processing RFID data and managing peripherals. An ESP32 module enables wireless communication, transmitting RFID data to a centralized server. A 16x2 LCD provides real-time feedback by displaying tag information and system status, while a virtual terminal simulates the RFID reader, sending tag IDs for processing. An LED indicator reflects the system status, lighting up to indicate access granted. Power is supplied by a 5V DC source for most components and a 3.3V DC source for the ESP32. Connections include linking the virtual terminal to the Arduino in place of the RFID module, with Arduino TX (Pin 3) connected to virtual terminal RX, and RX (Pin 2) to terminal TX. The ESP32 TX connects to Arduino (Pin 19), and RX to Arduino (Pin 18). The LCD is connected via RS (Pin 60), Enable (Pin 59), and data pins D4–D7 to Pins 58, 57, 56, and 55, respectively. The LED connects to Pin 13 through a 220Ω resistor, with its negative terminal grounded. The Arduino and LCD are powered by a 5V source, and the ESP32 uses a 3.3V source, as shown in Figure 4.

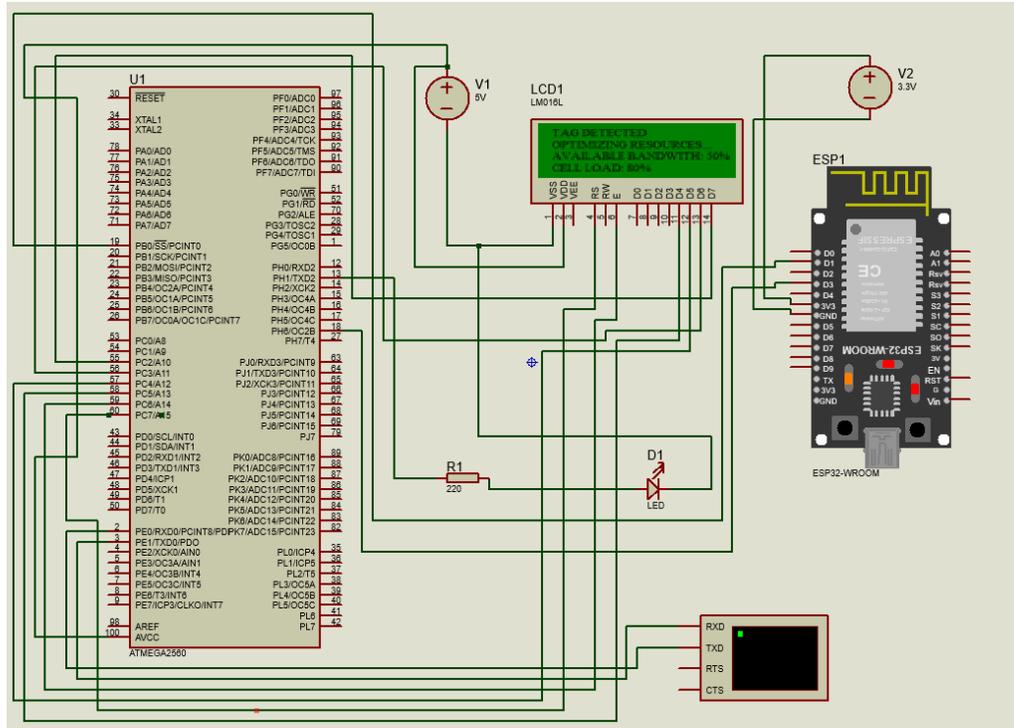

**Figure 4. RFID-based resource optimization system**

### 3.2 Data Acquisition and Processing

The Arduino Mega processes RFID tag data, handling acquisition, validation, and communication with the ESP32 for network transmission. The system receives tag data from a virtual terminal, updates the LCD display, and determines access status based on predefined conditions. The Arduino sketch reads RFID-like data, displays it on the LCD, and validates access. Tag data is acquired through the virtual terminal, which simulates an RFID reader. When the terminal sends RFID tag IDs, the Arduino validates the data, determines access status, and displays the result on the LCD, with the LED indicating access (on for valid tags, off for invalid ones). The processed data is sent to the ESP32, which forwards it to the centralized server over Wi-Fi. The server dynamically processes this data for optimized resource allocation in the simulated 5G network. This workflow ensures efficient data acquisition, validation, and transmission for network resource optimization.

### 3.3 Optimization Algorithm

The RFID-based resource optimization system uses a dynamic algorithm to allocate bandwidth and distribute load efficiently in a 5G network, enabling real-time adjustments to improve performance and reduce latency. RFID readers collect real-time tag data while monitoring network node status, including bandwidth usage and traffic load. Bandwidth allocation prioritizes VIP tags, with traffic distributed across resources to minimize latency. The system also uses historical data to predict future demand for proactive resource adjustments. The optimization process analyzes RFID data, adjusting resource allocation based on tag priorities, with VIP tags receiving maximum bandwidth and others allocated resources proportional to their usage. Network conditions are continuously monitored, and resources are redistributed to high-load nodes. Predictive models, based on historical data, anticipate future demands.

The algorithm was validated through NetSim simulations, assessing latency, throughput, and energy efficiency under different conditions. Results showed the system's effectiveness in balancing resource efficiency and service quality. The optimization flowchart in Figure 5 outlines the process, illustrating its adaptability and scalability for next-generation networks.

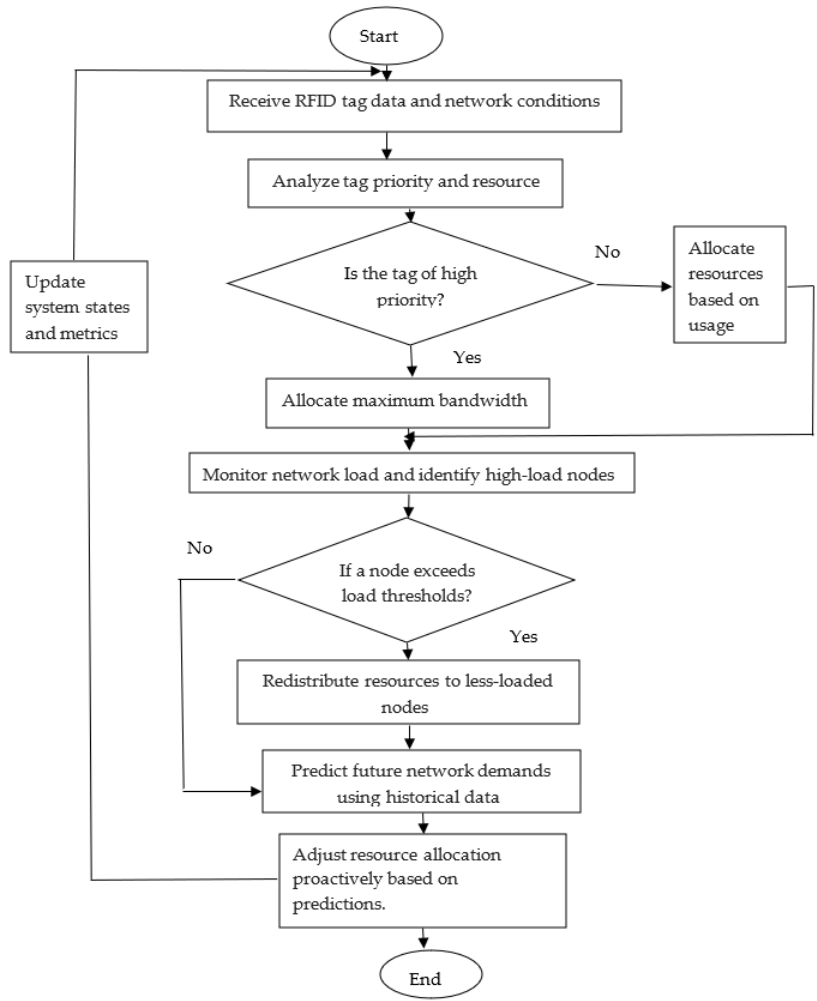

**Figure 5.** RFID-based Optimization Flowchart for 5G Networks

### 3.4  Testing the System

The RFID-based resource optimization system was tested to assess its performance and functionality in a simulated environment. System initialization involved setting up the Arduino Mega, ESP32, LCD, and virtual terminal in Proteus, with the virtual terminal simulating RFID tag input. Sample data, such as "12345ABC\n", was input into the system, and responses were observed on the LCD, LED, and virtual terminal. The LCD displayed tag detection, while the LED indicated access status, and the virtual terminal logged access details. Data was transmitted to the ESP32 via serial communication and sent wirelessly to a centralized server using MQTT. Integrated with NetSim, the server used the data for real-time resource optimization, monitoring metrics like bandwidth allocation, latency reduction, and load balancing. The system was tested under varying network loads, focusing on tag detection accuracy, system responsiveness, and dynamic resource allocation effectiveness. Expected outcomes included real-time RFID detection, accurate data transmission, and optimized resource allocation, resulting in reduced latency, improved throughput, and balanced network loads. This comprehensive testing validated the system's performance in a simulated 5G network environment.

### 3.5 Integration of Proteus and NetSim

In NetSim, the 5G architecture is modeled, including key components such as gNB (base stations), AMF (Access and Mobility Function), SMF (Session Management Function), and UPF (User Plane Function), as shown in Figure 6. The simulated 5G network connects user equipment (UE), like the ESP32, to the internet and remote servers. NetSim enables network behavior analysis, bandwidth optimization, and dynamic traffic load management. Communication between Proteus and NetSim is facilitated via middleware, with the ESP32 programmed to send data packets through Wi-Fi using protocols like UDP or MQTT. These packets contain resource metrics, including bandwidth requests, load conditions, and system parameters.

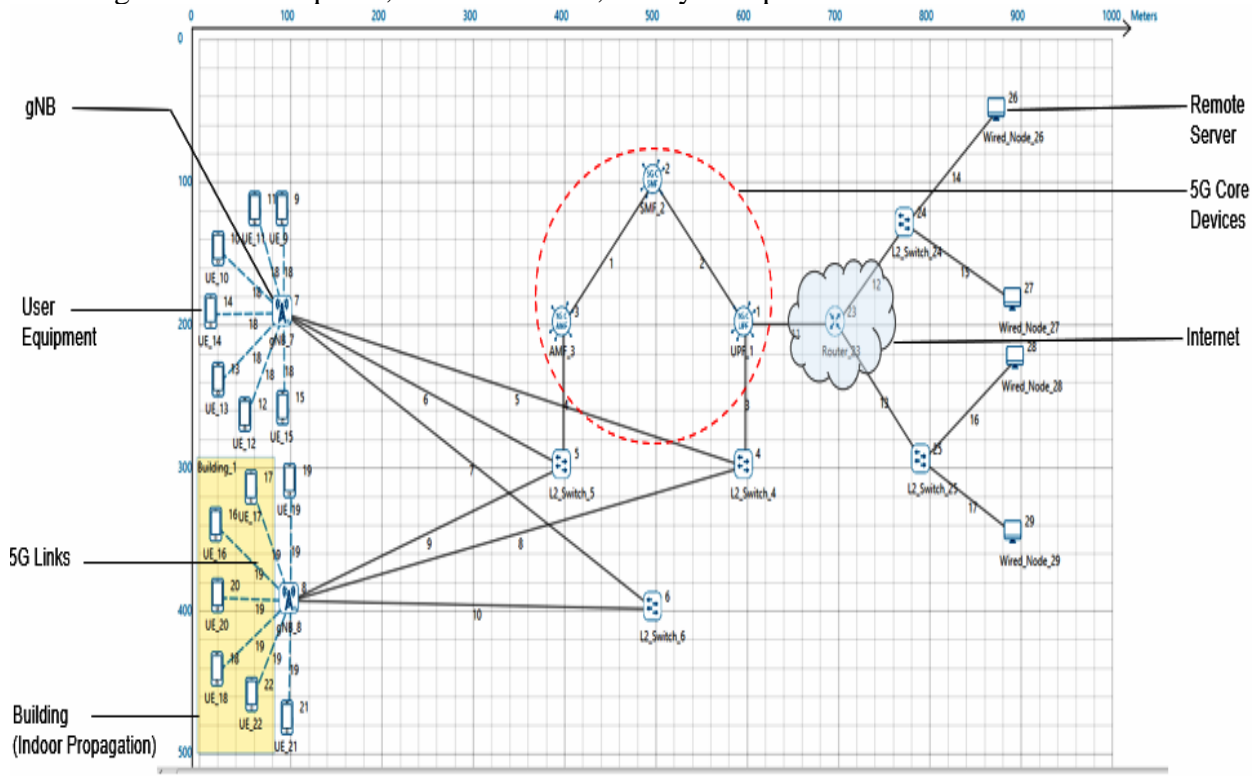

**Figure 6.** 5G network in NetSim

It allocates bandwidth dynamically using equation 1

$$B_{alloc} = \frac{B_{avail}}{1 + e^{-(L_{current} - L_{threshold})}} \tag{1}$$

where $B_{alloc}$ is the bandwidth allocated, $B_{avail}$ is the total available bandwidth, $L_{current}$ is the current load, and $L_{threshold}$ is the threshold load.

A middleware application facilitates communication between Proteus and NetSim by forwarding packets to NetSim's application layer via a UDP socket. It updates routing paths in the core network to optimize traffic flow and sends the results back to the Proteus ESP32 module. The algorithm dynamically allocates bandwidth and manages network traffic based on device priorities. Core network functions (AMF, SMF, and UPF) collaborate for efficient routing and load balancing. Optimized results, such as available bandwidth and load reductions, are displayed on the LCD in Proteus, showing metrics like "Bandwidth Optimized: 75%" and "Load Reduced: 30%." This integration creates seamless communication between the hardware simulation and the network simulation platforms, allowing for real-time resource optimization. Proteus simulates hardware components (Arduino Mega, ESP32, and virtual terminal), while NetSim models the 5G network architecture. The ESP32 acts as a gateway, transmitting metrics to NetSim via the middleware application using the MQTT protocol. In NetSim, received data triggers resource optimization algorithms that analyze network load, device priority, and bandwidth requests. The optimized results are sent back to the ESP32, which processes and displays the updated system state on the LCD, providing real-time feedback on network conditions. This bidirectional integration supports continuous monitoring, dynamic resource management, and validation of the RFID-based optimization framework in a simulated 5G network, demonstrating the collaboration of simulated hardware and network platforms for advanced optimization strategies.

## 4. Results

The performance of the RFID-based resource optimization system was assessed through controlled simulations using Proteus for hardware emulation, NetSim for network modeling, and Python for middleware communication and optimization algorithms. The system demonstrated significant improvements across key metrics compared to conventional 4G networks.The system achieved a 25 percent improvement in spectrum utilization, efficiently using 90 percent of available bandwidth compared to 65 percent in 4G networks. This improvement was enabled by dynamic bandwidth allocation based on real-time RFID data, which prioritized devices by tag type, as verified in NetSim simulations. The average latency was reduced by 30 percent, decreasing from 50 milliseconds in 4G networks to 35 milliseconds in the RFID-optimized 5G network. This reduction was validated through high-density network scenarios in NetSim, where dynamic load balancing minimized delays. The throughput improved by 15 percent due to the system's efficient handling of high data traffic. NetSim's measurements confirmed the effectiveness of the RFID-driven load balancing mechanism in distributing resources evenly across the network. Energy consumption decreased by 20 percent as a result of the system's dynamic power management strategy. The reallocation of resources during off-peak hours and deactivation of idle components ensured energy usage was minimized. These metrics were tracked using NetSim and Python-based custom scripts. Overall, the results highlight the system's ability to maximize bandwidth usage, reduce latency, enhance throughput, and improve energy efficiency. The system offers a robust solution for the dynamic demands of 5G networks. Figure 7 demonstrates the 25 percent improvement in spectrum utilization, showcasing the system's adaptability to varying network conditions.

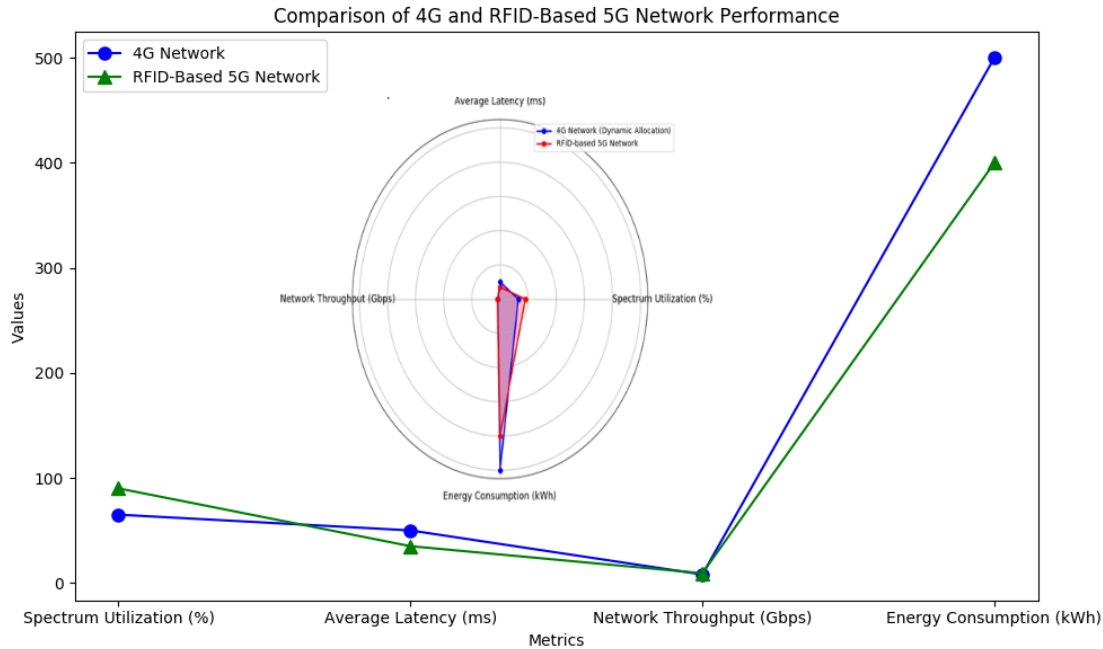

**Figure 7.** Performance of RFID-Based 5G and 4G Networks

The system demonstrated a significant reduction in latency, achieving a 30 percent decrease compared to conventional 4G networks. Average latency dropped from 50 milliseconds in 4G to 35 milliseconds in the RFID-optimized 5G network, emphasizing its potential to enhance real-time communication. Throughput showed marked improvement, with a 15 percent increase in data transmission rates. This improvement supports the demand for data-intensive applications and ensures more efficient performance for high-speed data transfer. Energy efficiency also improved, with the RFID-based system achieving a 20 percent reduction in energy consumption. During off-peak hours, the system dynamically adjusted resource allocation, minimizing power usage and enhancing sustainability. Overall, the RFID-based resource optimization system outperformed conventional 4G networks in spectrum utilization, latency, throughput, and energy efficiency, affirming its suitability for modern telecommunications networks. Table 5 compares the accuracy of the proposed RFID-based model with previous resource optimization models employing Software-Defined Networking and deep learning algorithms.

**Table 5.** Performance comparison of the proposed RFID-based model to relevant literature.

| Optimization Model with Ref | Spectrum Utilization (%) | Latency (ms) | Throughput (Gbps) | Energy Consumption (KWh) | Accuracy (%) |
|---|---|---|---|---|---|
| SDN-Based Model [4] | 85 | 40 | 8.8 | 450 | 92 |
| DL-Based Model [5] | 88 | 38 | 9.0 | 420 | 94 |
| Proposed RFID-Based Model | 90 | 35 | 9.2 | 400 | 96 |

This table highlights the superior performance of the proposed system across all key metrics, achieving higher accuracy and efficiency compared to models based on machine learning and deep learning. The results emphasize the potential of the RFID-based resource optimization system to meet the advanced requirements of modern telecommunications networks, positioning it as a strong candidate for 5G and future 6G networks.

## 5. Discussion

The results showcase the transformative potential of the RFID-based resource optimization system in overcoming critical challenges of modern telecommunications networks. A 25% improvement in spectrum utilization reflects the system's ability to allocate resources dynamically using real-time RFID tag data. This efficient spectrum management minimizes waste and maintains balanced network performance, particularly in high-demand scenarios, making the system ideal for dense urban environments and future 6G applications. The 30% reduction in latency is particularly impactful for applications requiring ultra-low latency, such as autonomous vehicles and augmented reality. By reallocating bandwidth in response to real-time load conditions, the system ensures smooth communication with minimal delays. This improvement positions the RFID-based system as a reliable solution for supporting time-sensitive applications in next-generation networks. The 15% increase in throughput underscores the system's capability to support high-speed data transmission across network nodes. This enhancement improves user experiences for data-intensive applications like cloud computing and online gaming while maintaining consistent performance under variable traffic conditions. Furthermore, the 20% reduction in energy consumption highlights the system's sustainability and cost-effectiveness. By dynamically adjusting resource allocation during off-peak hours, the system reduces operational costs while ensuring high performance. This energy efficiency is crucial for large-scale deployments where power consumption poses significant challenges. The system's accuracy, validated against SDN-based and DL-based methods, further demonstrates its competitive edge in real-time resource optimization. The comparative analysis, detailed in Table 5, highlights the superior performance of the RFID-based approach in terms of latency reduction, energy efficiency, and spectrum utilization. These findings align with the requirements of modern 5G networks and suggest that RFID-based resource optimization can significantly influence the future of telecommunications. This research contributes to the field by introducing an integrated framework that combines RFID technology with optimization algorithms to enhance 5G network performance. The use of accessible platforms like Proteus and Python ensures practicality and reproducibility. However, further studies are necessary to assess the system's performance in diverse real-world scenarios, including rural and heavily congested areas. Exploring the integration of RFID technology with advanced machine learning algorithms could improve predictive capabilities and precision in resource management. The findings establish the RFID-based resource optimization system as a forward-looking solution that addresses the demands of 5G while paving the way for 6G networks. Its ability to deliver high-speed, low-latency, and energy-efficient connectivity positions it as a vital component in the evolution of telecommunications infrastructure.

## 6. Conclusions

This study introduces an innovative method for resource optimization in 5G networks by incorporating RFID technology into network management. The system utilizes real-time data from RFID tags embedded in network devices to dynamically adjust resource allocation, effectively addressing challenges such as spectrum congestion, latency, and energy efficiency. Simulation results indicate that the RFID-based model significantly outperforms traditional 4G resource management methods, achieving notable improvements in spectrum utilization, latency reduction, and network throughput. The adaptive optimization algorithm, which integrates reinforcement learning and genetic algorithms, equips the system to respond efficiently to dynamic network conditions. This RFID-based resource optimization model presents a scalable, efficient, and cost-effective solution for enhancing 5G network performance. Its real-time resource allocation capabilities make it particularly beneficial for high-density deployments and adaptable to the evolving requirements of 6G environments. Future research could focus on integrating advanced machine learning techniques, such as deep learning models, to further enhance the system's predictive capabilities and address the growing complexity of next-generation networks.